\documentclass[12pt]{spieman}  
\usepackage{amsmath,amsfonts,amssymb}
\usepackage{graphicx}
\usepackage{setspace}
\usepackage{tocloft}

\title{PRIMA survey of the Virgo cluster}

\author[a,*]{Jacopo Fritz}
\author[b]{Maarten Baes}
\author[a]{Diego Alejandro Vasquez-Torres}
\author[c]{Karla Alejandra Cutiva-Alvarez }
\author[b,d]{Angelos Nersesian}
\author[e]{Viviana Casasola}
\author[a]{Eric F. Jiménez-Andrade}
\author[f]{Simone Bianchi}
\author[g]{Luca Cortese}
\author[b]{Ilse De Looze}
\author[h]{Frédéric Galliano}
\author[h]{Suzanne Madden}
\author[i]{Matthew W.L. Smith}
\author[j]{Manolis Xilouris}
\affil[a]{Instituto de Radioastronomía y Astrofísica, Universidad Nacional Autónoma de México, Morelia, Michoacán 58089, Mexico}
\affil[b]{Sterrenkundig Observatorium, Universiteit Gent, Krijgslaan 281 S9, B-9000 Gent, Belgium}
\affil[c]{Astronomy Department, University of Guanajuato, Apartado Postal 144, 36000 Guanajuato, Mexico}
\affil[d]{STAR Institute, Université de Liège, Quartier Agora, Allée du Six Août 19c, 4000 Liège, Belgium}
\affil[e]{INAF – Istituto di Radioastronomia, Via P. Gobetti 101, 40129 Bologna, Italy}
\affil[f]{INAF-Osservatorio Astrofisico di Arcetri, Largo E. Fermi, 5, 50125, Firenze, Italy}
\affil[g]{International Centre for Radio Astronomy Research, The University of Western Australia, 35 Stirling Highway, Crawley, Perth, WA 6009, Australia}
\affil[h]{Université Paris-Saclay, Université Paris Cité, CEA, CNRS, AIM, 91191, Gif-sur-Yvette, France}
\affil[i]{School of Physics \& Astronomy, Cardiff University, The Parade, Cardiff CF24 3AA, UK}
\affil[j]{Institute for Astronomy, Astrophysics, Space Applications \& Remote Sensing, National Observatory of Athens, P. Penteli, 15236 Athens, Greece}

\cftpagenumbersoff{figure}
\cftpagenumbersoff{table} 
\begin{document} 
\maketitle

\begin{abstract}
Clusters of galaxies are unique laboratories for investigating the dependence of galaxy evolution on their environment. The Herschel Virgo Cluster Survey (HeViCS) mapped the central $\sim$84 deg$^2$ region of the Virgo Cluster in five bands between 100 and 500 $\mu$m, which resulted in the first detailed view of cold dust in cluster galaxies. Major limitations of the HeViCS survey were the lack of data, or its limited availability, in the 20 to 80 $\mu$m range, and the quite low sensitivity of the Photodetector Array Camera and Spectrometer instrument, resulting in poor constraints on the warmer dust component. The PRIMAger instrument onboard PRIMA offers the capability to map a large portion of the Virgo Cluster—including regions beyond its virial radius—in hyperspectral and polarimetric bands from 25 to 265 $\mu$m, enabling a direct comparison with the area previously covered by HeViCS. By combining PRIMA and Herschel data with existing multi-wavelength photometry, it becomes possible to explore the connection between stellar and dust properties in a complete sample of cluster galaxies; to investigate environmental effects on the warm dust component within the Virgo Cluster; to map the magnetic field structure of the cold interstellar medium (ISM); to search for dust emission from the intra-cluster medium; and to study the ISM in background galaxies projected behind the cluster.
\end{abstract}


{\noindent \footnotesize\textbf{*}Jacopo Fritz,  \linkable{j.fritz@irya.unam.mx} }

\begin{spacing}{2}   

\section{Introduction}
\label{sect:intro} 
The interstellar medium (ISM) of galaxies residing in clusters is subject to a number of external phenomena caused both by the relatively high density of galaxies and by the presence of the very hot intracluster medium (ICM) permeating these structures. Such phenomena can be both hydrodynamical and gravitational in nature: among the first kind are ram pressure\cite{gunn72}, thermal evaporation\cite{cowie77}, and viscous stripping\cite{nulsen82} whereas tidal interactions\cite{merritt83}, strangulation/starvation\cite{larson80}, and harassment\cite{moore96} are gravitational. As the ISM provides the raw material for star formation, any change in its content, density, or distribution can possibly affect a galaxy's star-forming ability and efficiency, both with a positive (bursting) and negative (quenching) effects. In fact, although gas removal from galaxies in clusters has been observed for decades now, in some cases, compressive tidal interactions or gas stripping might initially induce localized starbursts before quenching takes over due to gas depletion. Furthermore, as stellar evolution is strictly related to the overall chemical enrichment, these processes link the evolution of individual galaxies to the chemical enrichment of the ICM. Material from star-forming regions is eventually recycled back into the cluster environment, contributing to this enrichment.

One important component of the ISM, even though minor with respect to its mass, is interstellar dust. Primarily heated by either the interstellar radiation field, by newly born stars in star forming regions, or by emission from gas accretion onto a supermassive black hole, dust re-emits at mid-infrared (MIR) and far-infrared (FIR) wavelengths dominating emission in galaxies within these frequency domains. As part of the ISM, dust is potentially affected by the cluster environment as well, plus it provides clues to characterize star formation in such a hostile environment\cite{boselli06,kenney15}. Dust properties such as size distribution and chemical composition are indirectly used to trace recent star formation activity, and they are affected by cluster-specific processes, including thermal sputtering in the ICM\cite{draine79} and shocks from ram pressure, which can fragment or alter dust grains (see e.g. Refs.\cite{tielens94} and\cite{vollmer08}). These  destruction mechanisms are not only indicators of the immediate physical conditions in the ICM, but also impact future star formation indirectly, by depleting the dust reservoir necessary for molecular gas shielding and cooling. MIR and FIR data thus provide an observational window into how dust evolves under cluster conditions and contributes to the chemical enrichment of both the ISM and the ICM.

In this context, the Virgo Cluster is of particular importance for several key reasons:\\
\underline{Proximity}: located at a distance of about 17 Mpc\cite{blakeslee09} , Virgo is the nearest large galaxy cluster to the Milky Way. This proximity allows for detailed observations across a wide range of wavelengths, providing the highest spatial resolution data on its galaxies and its ICM, and it is often viewed as a prototypical laboratory for studying galaxy evolution and interactions\cite{boselli06,ferrarese12} .\\
\underline{Diversity of Galaxies}: Virgo contains a rich variety of galaxy types, from massive ellipticals to smaller spirals and irregular galaxies, each exhibiting different ISM properties. This diversity makes Virgo an ideal target for studying how different galaxies interact with their environment, particularly how the cluster’s ICM affects their ISM and star formation.\\
\underline{Interactions and environmental effects}: Cluster galaxies are notoriously subject to a variety of effects that drive both their characteristics and their evolution. The ISM in particular, can be influenced by both gravitational and hydrodynamical interactions. The first, caused either by the interaction with other cluster members or by the gravitational potential of the whole cluster, will also act on the stellar component. The latter, on the other hand, are mainly due to the interaction with the ICM. These processes, which will eventually lead to the removal of a galaxy’s ISM, can be observed in detail, offering insights into how cluster environments affect gas content and eventually quench star formation.\\
\underline{Intracluster Medium Studies}: Virgo has a well-studied ICM, which plays a vital role in galaxy evolution. Studying the ICM’s interaction with the ISM of cluster galaxies is key into understanding processes like the ISM heating, the suppression or triggering of star formation, and the distribution of metals in both the galaxies and the ICM itself.\\
\underline{Benchmark for Galaxy Cluster Evolution}: as one of the nearest and best-studied clusters, Virgo serves as a benchmark for comparison with more distant ones. Insights gained from Virgo about ISM behavior, galaxy dynamics, and interactions between galaxies and the ICM, can be applied to understand clusters at higher redshifts.

This paper highlights the scientific potential of the PRIMA\cite{moullet23} Hyperspectral Imager (PHI; see Ref. \cite{meixner24}) and the PRIMA Polarimetric Imager (PPI; see, e.g., Ref. \cite{Yates25}) through the case study of spectrophotometric observations of the Virgo galaxy cluster. We discuss the feasibility and scientific scope of surveying an 84-deg$^2$ field—comparable in area to previous FIR surveys such as Herschel Virgo Cluster Survey (HeViCS)—using moderate integration times (e.g., 1 h/deg$^2$), demonstrating how such a program would enable key investigations into dust properties, ISM structure, and cluster environmental effects. This region encompasses the Virgo Cluster's virial radius, estimated at approximately 4 deg, \cite{planck_coll16}, and extends beyond it to include galaxies residing in outer substructures, such as the northwest (NW) cloud, west (W) cloud, and the southern extension\cite{binggeli85}, known to be part of the cluster.

\section{Virgo Multiwavelength Observations}
\begin{figure}
\centering
\includegraphics[width=0.75\linewidth]{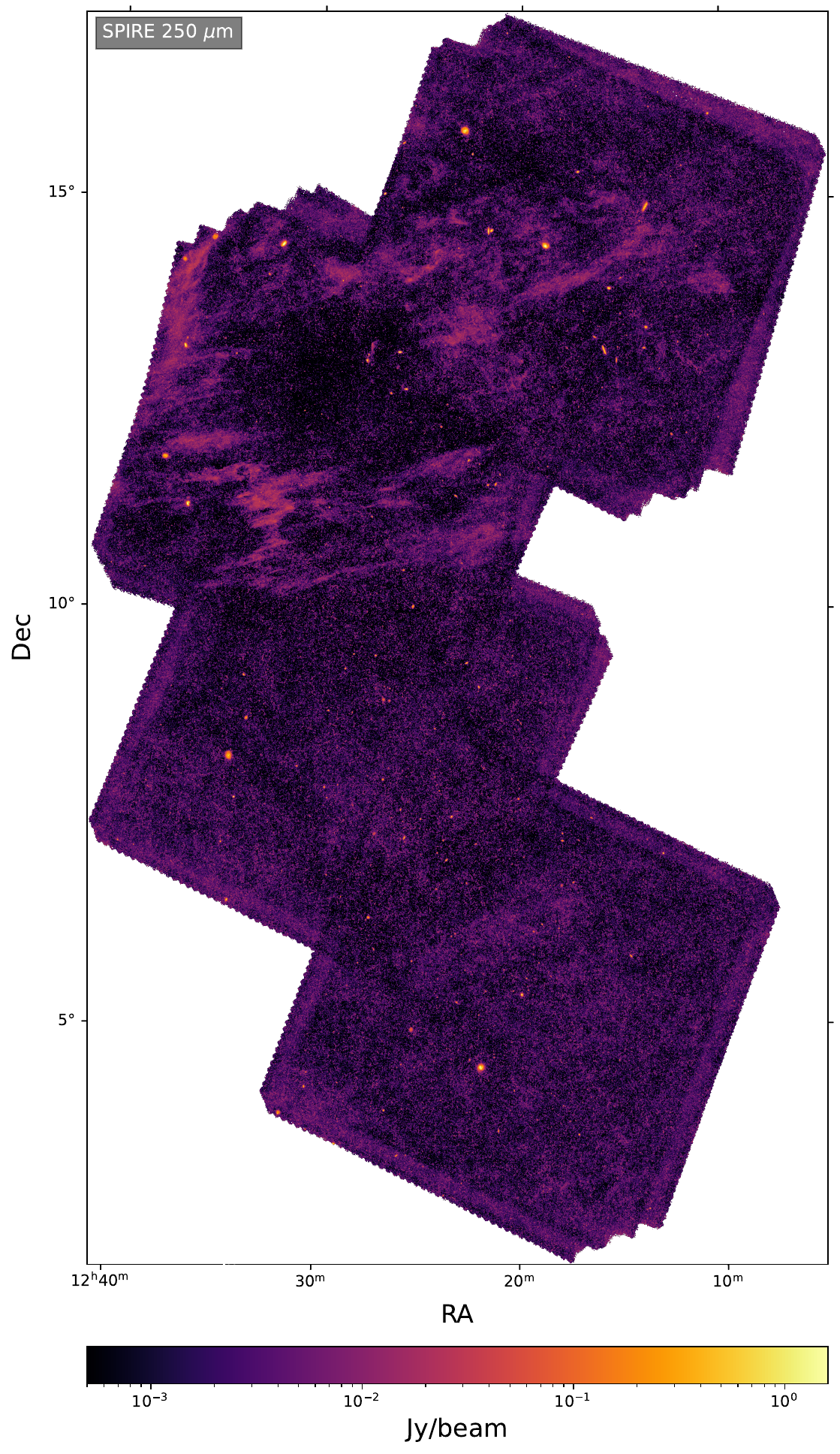}
\caption{Map of the Virgo cluster at 250$\mu$m taken by the SPIRE instrument onboard the Herschel satellite. The total coverage is about 84 square degrees.}
\label{fig:spire250}
\end{figure}

In recent years, the Virgo Cluster has been extensively surveyed across multiple wavelengths, providing unique insights into different components of the cluster, including its galaxies, the ICM, and the interactions among them. These efforts resulted in the collection of a wealth of data available to the astrophysical community, spanning the whole electromagnetic spectrum, from X-ray band\cite{bohringer94} to radio, with the ALFALFA survey\cite{giovanelli05} (but see also VIVA\cite{chung09}, ViCTORIA\cite{spasic24}, and VERTICO\cite{brown21}). The optical and near infrared (NIR) domains have been thoroughly covered by several surveys such as the Virgo Cluster Catalog (VCC\cite{binggeli85}), the Next Generation Virgo Cluster Survey (NGVCS\cite{ferrarese12}), and the NIR all-sky survey 2MASS\cite{skrutskie06}. In the MIR and FIR ranges, Virgo has been observed by IRAS\cite{neugebauer84} as a part of its all-sky survey, and was specifically targeted by the Herschel IR satellite with the Herschel Virgo Cluster Survey (HeViCS\cite{davies2010}) that observed galaxies at 100, 160, 250 350, and 500 $\mu$m, with a sub set of 143 cluster members also being observed by Photodetector Array Camera and Spectrometer (PACS) at 70 $\mu$m, in a series of follow-up observations\cite{davies19}. In Fig.~\ref{fig:spire250} we show the SPIRE 250 $\mu$m  map taken as a part of the HeViCS project, and which constitutes the base of the scientific case we present here.

The Herschel IR satellite mapped an area almost fully covering the Virgo Cluster with the HeViCS program in five  bands, from 100 to 500 $\mu$m (see Fig.~\ref{fig:spire250}). This wavelength coverage mainly allowed the analysis of the cold (i.e. $T \sim 15 \mbox{ to } 30$ K) dust component that mostly emits in the sub-millimeter range and is crucial for understanding the bulk mass of the dust in a galaxy. Herschel's observations missed --by their very nature-- the warmer component of dust, whose heating sources are supposed to mainly come from star formation or nuclear activity, and whose emission peaks in the MIR. 

The Virgo cluster has only very sparse observations in this range, with the exception of WISE (at $\sim 22$ $\mu$m) and IRAS (at 25 and 60 $\mu$m) data, which are both characterized by quite low resolutions and sensitivities. PRIMA is a strong leap forward with respect to the first aspect in particular: with an average beam size of $4^{\prime\prime}$ in the 24-45 $\mu$m range, it is three  times better than WISE (band 4), and will provide a resolution comparable to or better than SPIRE at 250 $\mu$m (FWHM$\simeq 18^{\prime\prime}$) across most of the observable range it will offer. The benefits are even more striking when considering an IRAS beam of $\sim3^\prime$ at 100 $\mu$m.

A survey of the Virgo Cluster covering the same area imaged by the PACS and SPIRE instruments onboard Herschel--$\sim 84$deg$^2$--can be carried out with the PRIMAger instrument using an integration time of 1 h/deg$^2$. Based on PRIMAger's expected characteristics and specifications\footnote{See: \url{ https://prima.ipac.caltech.edu/page/instruments}}, such a survey would achieve the sensitivities listed in Table~\ref{tab:sensitivity}. The values correspond to 5$\sigma$ detection in background subtracted maps.

\begin{table}[ht]
\caption{Estimated sensitivities for 1 h / deg$^2$ observation.} 
\label{tab:sensitivity}
\begin{center}       
\begin{tabular}{|lr|c|c|c|c|c|c|} 
\hline
                      &    &    PHI1    &    PHI2    & PPI1 & PPI3 & PPI3 & PPI4 \\
\hline
Point Source Total & [mJy]                 & 3.73--4.93 & 6.96--12.97 & 5.60 &  8.10 & 10.72 & 14.51 \\ 
\hline
Point Source Polarized & [mJy]             &     --     &     --      & 7.91 & 11.45 & 14.70 & 20.52 \\
\hline
Extended Source Total & [MJy sr$^{-1}$]    & 5.19--2.09 & 2.34--1.83  & 1.45 &  1.08 &  0.79 &  0.57 \\  
\hline
Extended Source Polarized & [MJy sr$^{-1}$]&     --     &     --      & 2.06 &  1.49 &  1.11 &  0.79 \\
\hline
\end{tabular}
\end{center}
\end{table} 

The HeViCS map provides a valuable reference for FIR studies of the Virgo Cluster, but currently lacks homogeneous coverage in MIR range ($\sim 25$ to $\sim 80 \mu$m). Observations with the PRIMA Hyperspectral Imager can fill this gap, enabling uniform spectral mapping across this critical range. In addition, the PRIMA Polarimetric Imager opens the possibility to trace polarized dust emission in the $\sim 90$ to $\sim 230 \mu$m range, offering new insights into the magnetic field structure of the cold ISM.

We first discuss the improvements on Spectral Energy Distribution (SED) fitting due to the presence of these new data, particularly in the MIR range, and the repercussions on the results (Sec. ~\ref{sec:globalSED}). In Sec.~\ref{sec:extendedsources}, we discuss a number of cases to be addressed exploiting spatially resolved data, such as the expected detections of extended sources, dust depletion mechanisms, the possible detection of dust in the ICM, and the use of PRIMA data to disentangling dust from non-thermal emission in active galactic nuclei (AGN). Relying on archival data for sensitivity calculations and sources detection, in Sec.~\ref{sec:background} we present an estimate of the number of point sources expected to be detected in such a survey, the majority of which are background objects beyond Virgo, while Sec.~\ref{sec:magnetic} explores the potential of PRIMA's polarimetric observations, assessing their effectiveness for both spatially resolved and point-like sources.

\section{Global SED: Novel Insights from PRIMA Data}
\label{sec:globalSED}
The 24 to 90 $\mu$m range, which is the wavelength span covered by the Hyperspectral Imager, primarily captures warm ($T \sim 30$ to 120 K) dust heated by either of the aforementioned processes or the emission by the smallest, stochastically heated grains. This dust typically resides in {\sc Hii} regions or is mixed with the ISM, providing clues to the more active processes in galaxies. Emission in the $\sim 60-90 \mu$m regime is particularly interesting as it lies in between the peak emission of warm and cold dust, hence providing critical information about intermediate-temperature dust, and helps bridging the gap between hot star-forming regions and the more diffuse, cold dust in the ISM. 
 
Galaxies' properties, such as stellar mass, star formation history, and dust content, are typically derived by fitting their observed spectral SEDs with theoretical models. However, these results often suffer from degeneracies, where multiple combinations of model parameters can produce equally good fits to the data,  yielding (sometimes significantly) different physical interpretations. The absence of data in specific wavelength ranges exacerbates this issue. To highlight the advantages of incorporating mid-infrared coverage into the observed SED for this purpose, we applied two different dust models to reproduce the panchromatic emission of a typical Virgo spiral galaxy.

\begin{figure}
    \includegraphics[width=1\linewidth]{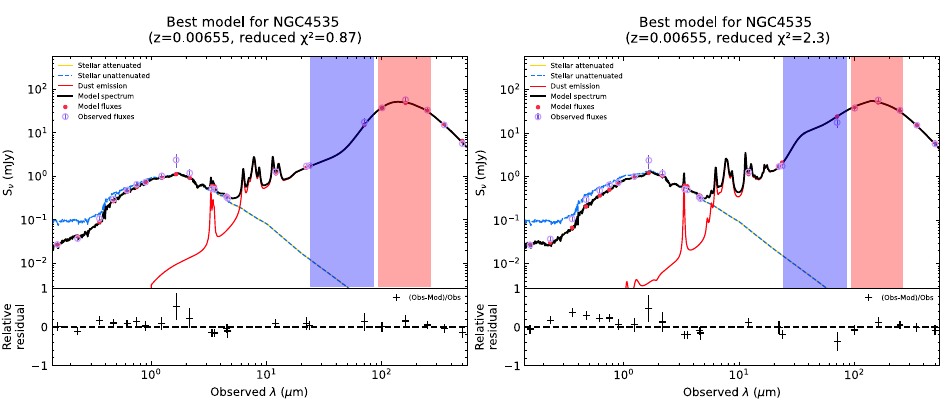}
\caption{Example of multi-component fitting of the Virgo spiral NGC4535. In the left-hand plot is the fit performed with the THEMIS dust model. On the right-hand plot, we show an alternative, equally acceptable fit, obtained with the \cite{draine07} dust model. The differences between the two models are maximal within the spectral range covered by the Hyperspectral cam, highlighted by the blue rectangle, while the red one covers the spectral range of the polarimetric imager.}
\label{fig:cigale_example}
\end{figure}

On the left hand panel of Fig.~\ref{fig:cigale_example}, we show the observed SED of NGC4535, a large spiral galaxy in Virgo, and its model obtained with the multi-wavelength fitting code CIGALE\cite{noll09} , as done by Ref. \cite{nersesian19} adopting the THEMIS\cite{jones17} dust model, following the prescriptions of the DustPedia\cite{davies17} project (see the DustPedia repository: \url{http://dustpedia.astro.noa.gr}). The spectral region sampled by the hyperspectral camera has not been systematically observed until now, whereas the polarimetry imager covers a wavelength range in the domain of the HAWC+ instrument onboard SOFIA, which is not operating anymore, furthermore adding spectroscopic capabilities, and both are expected to yeld far better sensitivities as compared with PACS. In the right-hand panel of the same figure, we attempted a fit to the same object and with the same fitting tool, but using different dust properties, namely those of Ref. \cite{draine07} . The use of two different dust models, yields different parameters for both the stellar populations and the dust properties. Fitting for which the adopted dust model is the one of Ref. \cite{draine07} resulted in a stellar mass of $7.20\times 10^{10}$ M$_\odot$, star formation rate of 3.8 M$_\odot$ year$^{-1}$, and a dust mass of $8.70\times 10^{10}$ M$_\odot$, whereas models assuming the THEMIS composition gave $3.71\times 10^{10}$ M$_\odot$, 3.6 M$_\odot$ year$^{-1}$, and $3.62\times 10^{10}$ M$_\odot$ for those quantities, respectively. Although both models provide a good fit to the cold dust emission (beyond $\sim100\ \mu$m), this somewhat extreme case highlights where divergences in the infrared emission tend to arise.

The comparison between the two models highlights the crucial role of PHI’s spectral coverage in distinguishing among different dust models, which otherwise reproduce the remaining portion of the IR SED equally well. Although the fit using the Draine dust model results in a higher $\chi^2$ value compared with that obtained with THEMIS, the complexity of dust emission modeling and the number of free parameters involved justify considering both fits as statistically acceptable (see also Ref. \cite{andrae10}). Nevertheless, the physical dust properties inferred from the two models differ significantly.

Properties such as dust temperature, size distribution, and composition can also be derived through SED fitting using detailed dust modeling. MIR data, provided by the Hyperspectral Imager, are essential in this context, as their absence makes it challenging to constrain these characteristics using only FIR and submillimeter (submm) data, as illustrated with the SED of NGC 4535. The modeling of an SED without these kind of data would lead to degeneracies in the derived properties, was also pointed out by Ref. \cite{galliano18b} (see their Fig. 3): without observations in the $\sim 30$ to $\sim 80$ $\mu$m range, the same data can be reproduced by both a combination of small and very small grains emission, or by a distribution of starlight intensity models as well. 

Pappalardo et al.\cite{pappalardo16} compared different fitting tools (namely MAGPHYS\cite{dacunha08} and CIGALE), which share a roughly similar energy balance approach to SED fitting, and highlighted how the absence of observed constraints in this range can result in very different model SEDs and, hence, in the derivation of different dust properties (see their Fig.14 clearly showing this effect).

\section{Spatially Resolved Dust Properties}
\label{sec:extendedsources}

When recovering the properties of dust in galaxies, any analysis based on integrated data will be biased toward the most luminous component, and will therefore provide only a partial view of the dust content. Spatially resolved analysis performed for each resolution element of a map, is an obvious solution to at least mitigate this issue, as it allows to disentangle the contribution from different regions of a galaxy, which can potentially display highly varying levels of physical conditions such as star formation, dust opacity, and metallicity. These characteristics play a fundamental role in shaping dust properties in the ISM\cite{bianchi22,casasola17,casasola22} . Spatially-resolved analysis exploiting IR or even multi-wavelength data has been performed increasingly often in the last years, especially in nearby galaxies thanks to {\it Herschel}'s improved resolution and spectral coverage over the previous IR probes \cite{smith10,viaene14} (but see also Ref. \cite{chastenet24} for spatially resolved fitting in the MIR-FIR). Dust emission modeling to interpret IRAS data was traditionally performed adopting a single temperature black body emission with a wavelength-dependent dust emissivity, whereas the quality and wavelength range of {\it Sptizer} and {\it Herschel} data pushed the use of more complex and physically motivated models\cite{chastanet17} . Still, the so-called modified black body emission was widely used for the interpretation of PACS and SPIRE data\cite{bianchi13}.

Smith et al.\cite{smith10} performed such kind of spatially-resolved analysis on three galaxies in Virgo, and noticed how a single temperature was only able to reproduce the IR SED if data below 100 $\mu$m were not included: the MIPS $70-\mu$m emission was constantly underestimated by the adopted model. Most pixels were found to present a significant flux excess at 70 $\mu$m compared with the fit, strongly indicative of the presence of a warmer dust component. While a two-component model would improve the fit, the warm component remains poorly constrained, as a wide range of temperatures yields statistically similar results due to the limited data on the short-wavelength side of the SED. The addition of MIR data from the PRIMA facility is expected to significantly improve constraints on the mass, temperature, and properties of the warmest dust components (see also Fig.~\ref{fig:cigale_example}). Moreover, access to the full MIR-to-FIR range enables the application of more physically motivated dust emission models (e.g., see Ref. \cite{galliano18a}).

\subsection{Extended Sources in Virgo} \label{ssec:largegalaxies}

An estimate of the capabilities of PRIMA on extended sources in Virgo, can be derived from the existing literature catalogs. To calculate the average surface brightness values at the relevant wavelengths, we exploit the catalog of Herschel fluxes of VCC galaxies presented in Ref. \cite{auld13} , together with the Extended Virgo Cluster Catalog (EVCC) presented by Ref. \cite{kim14} , where optical radius values are given. This allows to derive an estimate of the average surface brightness for Virgo cluster members. Of the 750 sources that are located in our field, we report 589 that are detected with PACS and 616 with SPIRE at 250 $\mu$m, for which we also have a measure of the radius from the catalog of \cite{kim14} . In Table~\ref{tab:ext_detect} we report a summary of the detections expected both in total intensity and in polarized light: for the latter, the estimated fluxes are calculated assuming a conservative assumption of a 1\% polarization. Estimates for the PPI1, PPI3 and PPI4 bands are obtained from PACS 100 and 160 and SPIRE 250 $\mu$m fluxes, respectively. For the two bands of the hyperspectral imager, PHI1 and PHI2, we rely on the IRAS Cataloged Galaxies and Quasars (available on the IRSA/IPAC website), which provides radius and flux density data for 85 Virgo sources at 25 and 60 $\mu$m.

\begin{figure}
    \includegraphics[width=1.01\linewidth]{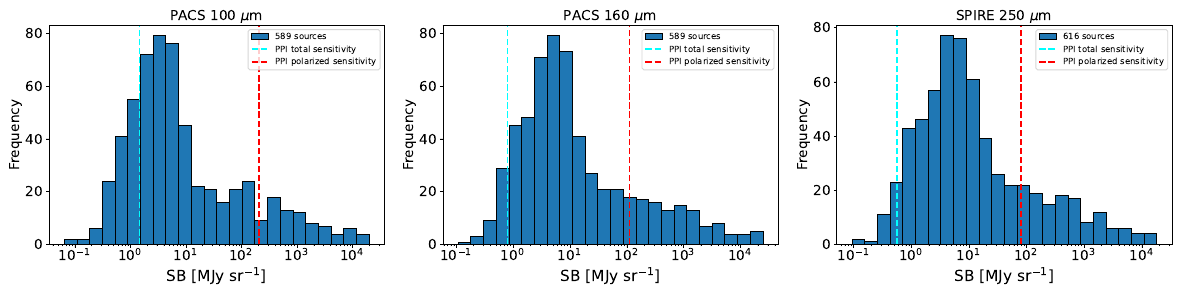}
    \caption{Surface brightness distributions of sources in PACS 100, 160 and SPIRE 250-$\mu$m bands. The vertical lines correspond to expected 5$\sigma$ detections for total (cyan) and polarized (red) intensity, for which we have assumed a 1\% polarization fraction.}
    \label{fig:ext_detect}
\end{figure}

In Fig.\ref{fig:ext_detect} we show the surface brightness distribution, in three polarimetry imager bands, of sources in the Virgo field we plan to observe.

\begin{table}[ht]
\caption{Estimated number of extended sources detection. The wavelengths values refer to the central wavelength of the instrument used for the reference flux (see text). Note that the total number of sources changes and is 85 for the PHI instrument and 750 for the PPI (see text for details).} 
\label{tab:ext_detect}
\begin{center}       
\begin{tabular}{|l|r|r|r|r|r|} 
\hline
              & PHI1  & PHI2  & PPI1  & PPI3 &  PPI4  \\
\hline
Wavelength ($\mu$m) & 25 & 60 & 100 & 160 & 250 \\
\hline
Total Power   &  53   & 85    & 462   & 552  &  592   \\    
\hline
Polarized     &  --   & --    &  80   & 113  &  125   \\    
\hline
\end{tabular}
\end{center}
\end{table} 

\subsection{Dust Depletion}
Different heating mechanisms and/or dust conditions, can be analyzed by exploiting a combination of MIR/FIR colors as done e.g. in Refs. \cite{draine07,boselli10} and Refs. \cite{cortese14,smith19,nersesian19} . Cluster galaxies are subject to environmental-specific phenomena affecting both the stellar and the interstellar medium (ISM) components. Specifically, hydrodynamical mechanisms such as ram pressure stripping\cite{gunn72}, thermal evaporation\cite{cowie77}, and viscous stripping\cite{nulsen82}, are expected to be quite efficient in affecting gas and dust properties. Potentially very useful control samples to compare Virgo galaxies with, are the SINGS\cite{kennicutt03} and KINGFISH\cite{kennicutt11} samples, with data available in a similar spectral range. DustPedia\cite{davies17} galaxies have already been used as a control sample to compare between different environments in\cite{davies19} , and can be used to build an even more complete control sample. Controlling by morphology and stellar mass, differences in the respective SEDs can be attributed to the effect of the local environment. 

Direct evidence of dust stripping in spiral galaxies of Virgo was found from the analysis of data at both optical\cite{abramson14} and FIR\cite{cortese10} wavelengths. In an analysis of the Hubble Space Telescope images of two Virgo galaxies displaying clear ram pressure stripping signature \cite{abramson14} found evidence for extra-planar dust clouds out to distances of 1.5 kpc from the galaxy disk, and similar results were found by Ref. \cite{longobardi20b} as well. On the other hand, the work of Ref. \cite{cortese10} exploiting Herschel data found a correlation between the ratio of the submillimetre-to-optical size and the {\sc Hi}-deficiency of the galaxies, indicating that galaxies that show evidences of substantial gas deficiency also present clear signs of dust stripping. The fact that these signatures manifest as truncation in the (cold) dust disk is a strong indication that ram pressure is the dominant mechanism. Data from PRIMA can extend this analysis by targeting warmer dust components, which may be more closely associated with the molecular gas. 

On the other hand, possible effects of other mechanisms such as dust sputtering due the presence of a hot gas, as firstly theoretically studied by Ref. \cite{draine79}, have been proposed but are more difficult to detect\cite{popescu00}. Although direct evidence for dust destruction via thermal evaporation is difficult to observe (because it results in a lack of dust rather than a visible feature; see e.g., Ref. \cite{clemens10}), these studies suggest that thermal sputtering is a powerful mechanism for dust destruction in the harsh cluster environment. 

\subsection{Dust in the ICM}
Bianchi et al.\cite{bianchi17} have thoroughly looked for diffuse emission from dust located within the ICM, and found that both the Cosmic Microwave Background and the galactic cirrus emission constitute a major factor of contamination, severely hampering ICM dust detection in the FIR and submm. Nevertheless, such diffuse dust component has been detected in absorption by Ref. \cite{longobardi20a} , in the innermost regions of the cluster, amounting to roughly 2.5$\times 10^9$ M$_\odot$. In principle, dust could be detected at hyperspectral camera wavelengths if its temperature was high enough with respect to the one of the Milky Way. Different heating mechanisms other than radiative, such as Coulomb interactions\cite{dwek92} or heating by photoelectric effect\cite{bakes94}, might be at play and can be efficient enough in a cluster environment.

\subsection{AGN in Virgo}
\label{ssec:AGN}
The Virgo Cluster is home to a few galaxies that exhibit clear signs of nuclear activity. Among early-type galaxies, notable examples displaying radio synchrotron emission are M49, M84, M87, and M89. As early-type galaxies are generally dust-poor, emission in the MIR and FIR regimes likely originates from a combination of synchrotron radiation and, potentially, dust emission as well. Reference \cite{baes10} modeled the whole spatially-resolved IR SED of M87, employing a synchrotron model to investigate the possible contribution of dust, hence setting an upper limit to the mass of a possible extended component. 

Dust in early-type galaxies, particularly in the cluster environment, is subject to destructive processes such as thermal sputtering by the hot ICM, shocks, and tidal interactions. Detecting dust in such environments not only provides constraints on its survival timescale under the extreme physical conditions characterizing in these galaxies\cite{clemens10}, but can also be used to estimate its production rate.

A spatially resolved analysis of the IR SED could provide valuable insight into the presence and distribution of dust in these systems. Such an approach would enable the estimation --or the determination of an upper limit-- of the total dust mass within these galaxies. Polarimetric data could further enhance this analysis by isolating the synchrotron component, allowing for a more accurate characterization of any residual dust emission.

\section{The Cosmic Volume Behind the Virgo Cluster} \label{sec:background}
In addition to the Virgo Cluster galaxies, an 84-deg$^2$ survey of this field can also detect numerous background sources at these wavelengths. The identification of these sources enables the exploration of additional scientific questions beyond the cluster itself. These background galaxies provide a window into the more distant universe, enabling studies of star formation, dust properties, and cosmic evolution at a variety of redshifts. Such data comes basically ``for free", and their exploitation has already provided interesting results on the Virgo field itself (see Refs. \cite{pappalardo15} and \cite{pappalardo16} ). 

The most straightforward way to estimate the number and types of background sources that PRIMA is likely to detect under the observing conditions outlined in this work is to leverage existing archival data and catalogs, such as those from {\it Herschel}, {\it IRAS}, or {\it WISE}. These datasets can offer a statistical prediction of the source counts at similar wavelengths and depths, helping to quantify PRIMA's contribution to our understanding of the cosmic volume beyond the Virgo Cluster.

A catalog of point sources detected within this area in Herschel bands, was presented in Ref. \cite{pappalardo15} . A total of 52,020 objects were detected at 250 $\mu$m (FWHM$\simeq 18^{\prime\prime}$), a band covering a very similar range to the polarimeter PP4 band, centered at 235 $\mu$m. 

\begin{figure}[!t]
    \centering
    \includegraphics[width=0.75\linewidth]{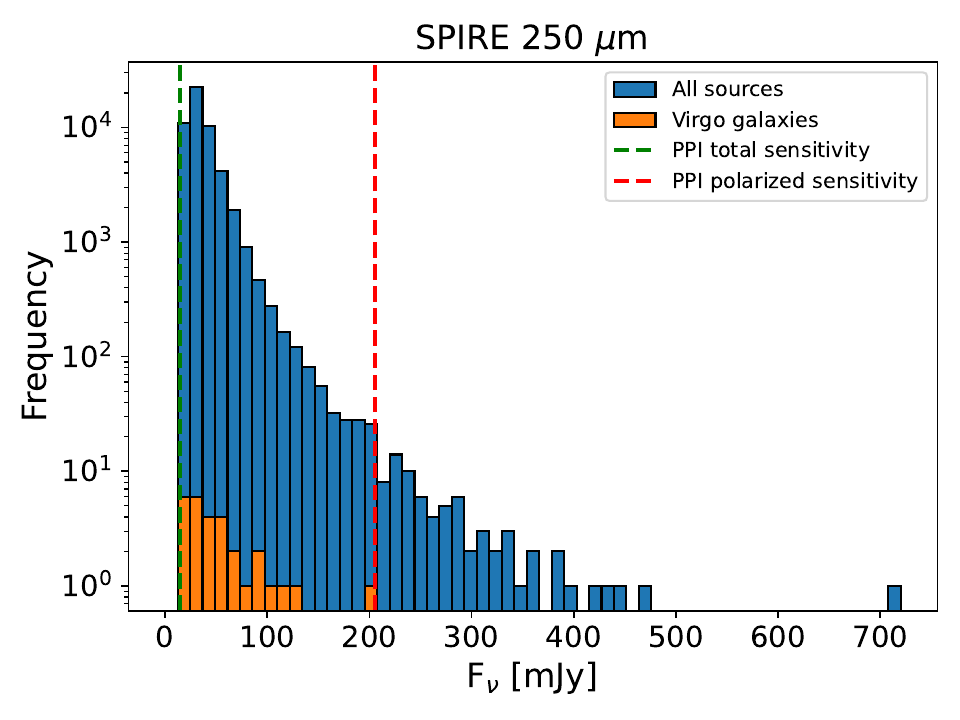}
    \caption{Source counts detected at 250 $\mu$m by SPIRE on {\it Herschel}, with a flux density limit of 20 mJy. The green and red lines correspond to the 5$\sigma$ detection limits for sources in total and polarized intensity, respectively.}
    \label{fig:250dist}
\end{figure}

In Fig.\ref{fig:250dist} we show the flux density distribution of SPIRE 250 sources detected within the Virgo field. With a 1-h observation per square degree, a 5$\sigma$ detection limit of 14.5 mJy for point sources can be achieved in total intensity (see Table~\ref{tab:sensitivity}), which should hence allow the detection of all of these sources (and likely many more) in the PP4 band, within a beam of $\sim 24^{\prime\prime}$. When applying a cut in redshift of 0.01, $\sim 30$ detectable point sources are found to (likely) belong to Virgo. Assuming a modified blackbody emission with $T=25$ K and an emissivity index $\beta =2$, typical of cold dust in galaxies, such sources would also be detectable in the other three bands of the polarimetry imager, at least up to z$\simeq 0.4$. However, at higher redshifts, the flux in band 1 progressively falls below the detection limit.

The 5$\sigma$ detection limit in polarized light is $\sim 20.5$ mJy which, under the conservative assumption of 1\% polarization—implies that only sources brighter than 2.05 Jy would be detected. No such objects are present in this field. Instead, if we consider a polarization fraction of 10\%, which is a common value found in AGN, and that corresponds to a limit of 0.20 Jy (red line in Fig.~\ref{fig:250dist}), we found that up to 74 objects can be detected. Polarization in AGN can arise from different emission mechanisms: relativistic jets emit synchrotron emission that can be studied across a very large frequency range, as done in M87 by Ref. \cite{baes10} , and that can reach polarization levels up to 30\% at radio frequencies\cite{Begelman84}. At FIR wavelengths, the synchrotron polarized fraction may decrease due to depolarizing effects and mixing with dust non-polarized emission, and measures at these wavelengths are currently basically missing.

On the other hand, the dusty torus in AGN, whose emission peaks in the 20–50 µm range, can be another significant source of polarized light. This has been studied in detail by \cite{marin17} using radiative transfer simulations, which predict average polarization levels of $\sim 1$\%, with values reaching up to 10 times higher in specific configurations, which can be detected in the highest redshift Quasi Stellar Objects. Using the catalog provided by Ref. \cite{cavuoti14}, we found $\sim 37500$ sources classified as AGN that are located within the targeted area.

The study of polarization in AGN at these wavelengths, even though not in a spatially resolved fashion, is potentially interesting for a number of reasons. In these objects, multiple emission mechanisms generally contribute to the total emission, such as synchrotron radiation from relativistic jets and thermal emission from the dusty torus, and the analysis of polarization can help disentangle these components. Synchrotron radiation is highly polarized, whereas thermal emission from dust typically shows lower polarization, such that it should be possible to separate and study these different components in more detail, by analyzing their polarized SEDs. 

In the case of distant AGNs that are not spatially resolved, polarimetric SEDs could provide hints on the dust grain sizes and compositions. The behavior of polarization across different wavelengths can indeed indicate specific processes related to dust grain alignment and the lifecycle of grains within interstellar environments. For instance, changes in the polarization degree as a function of wavelength can suggest specific grain alignment with the magnetic field or grain growth and destruction processes\cite{jones96} . The work carried out by Ref. \cite{lazarian07} has shown that changes in polarization intensity and angle at specific wavelengths often result from the alignment of dust grains with magnetic fields, influenced by radiative torques (RATs) and mechanical interactions. The same authors suggest that larger grains align more efficiently with magnetic fields, and shifts in polarization can signify alignment variations with respect to grain size and shape, especially in response to different radiation intensities. When the polarization decreases, it can indicate disruptions in grain alignment, potentially due to increased turbulence or gas drag, which disrupts such alignment. Alternatively, an increase in polarization might suggest more stable alignment due to radiation, or even the growth of grains, as larger grains are generally better aligned than smaller ones (see e.g. Refs. \cite{andersson15} and \cite{jones13}). 



\section{Magnetic Fields Characterization} 
\label{sec:magnetic}
Given Virgo's proximity, many of its galaxies exhibit relatively large angular sizes, often several times larger than PRIMA's expected beam at the longest wavelengths. Table~\ref{tab:spiral_sizes} reports data for the 13 largest spiral galaxies that are located within the PRIViCS field, for which we have calculated the expected surface brightness using PACS and SPIRE fluxes. To measure galaxies' size in the IR, we have visually inspected HeViCS 250 $\mu$m maps for the 13 galaxies, and we have used the value of their semi-axes to calculate the average surface brightness. With spatial resolutions of $\sim$ 0.8, 14.8, 20.2, and 27.6$^{\prime\prime}$ in the four polarimetric bands offered by PRIMA, and assuming a 1\% polarization fraction, none of the large spiral galaxies contained in the proposed field is likely to be detected in polarized flux with our observing strategy.

Hence, observing these galaxies in polarized light would require dedicated observations with significantly longer integration times than those adopted in the standard survey strategy. For instance, targeted exposures of $\sim$ 4 h per galaxy would provide the sensitivity needed to detect polarized flux in all of these targets in both the PP3 and PP4 bands, and in five of them also in PP1 and PP2. The latter band is particularly promising given the expected higher fluxes and better instrument sensitivity. This perspective outlines the observational conditions necessary to access the polarization properties of dust and probe magnetic field structures in these systems.


\begin{table}[ht]
\caption{Spiral galaxies with a diameter larger than 2 arcmin in the K$_s$ band (data retrieved from the NASA  Extragalactic Database). Objects marked with $^1$ and $^2$ are classified as galaxy-pair, with the latter pair being unresolved at 250 $\mu$m. $1a$ is an edge-on galaxy. We also report the average surface brightness in {\it Herschel} PACS (100 and 160 $\mu$m) and SPIRE (250 $\mu$m), in units of MJy sr$^{-1}$, and the major and minor axes of the SPIRE 250 $\mu$m image. } 
\label{tab:spiral_sizes}
\begin{center}       
\begin{tabular}{|l|l|l|c|r|r|r|r|r|} 
\hline
VCC         & Other ID & Type  & D$_K$ [$^{\prime\prime}$] & a$_{250}$  [$^{\prime\prime}$] & b$_{250}$ [$^{\prime\prime}$] & SB$_{100}$ & SB$_{160}$ & SB$_{250}$  \\
\hline
0092        &   M98    & SABab &   626.6  &    256  &   94   &  14.67  &  22.48 &  15.83  \\  
\hline          
0167        & NGC 4216 & SABb  &   635.6  &    226  &   74   &  12.42  &  22.43 &  17.67  \\  
\hline          
0307        &    M99   &  SAc  &   408.6  &    220  &  170   &  39.83  &  45.18 &  24.39  \\  
\hline          
0483$^1$    & NGC 4298 & SAc   &   150.1  &    106  &   78   &  22.99  &  32.60 &  20.79  \\  
\hline          
0497$^{1a}$ & NGC 4302 & Sc    &   429.0  &    194  &   64   &  17.69  &  28.81 &  21.61  \\  
\hline          
0508        &    M61   & SABbc &   399.2  &    210  &  190   &  33.22  &  36.27 &  18.75  \\  
\hline          
0596        &   M100   & SABbc &   651.8  &    210  &  200   &  29.13  &  36.31 &  22.40  \\  
\hline          
1401        &    M88   &  SAb  &   461.8  &    190  &  140   &  38.71  &  49.38 &  30.73  \\  
\hline          
1555        & NGC 4535 & SABc  &   410.2  &    220  &  170   &  13.57  &  18.84 &  12.59  \\  
\hline          
1673$^2$    & NGC 4567 & SAbc  &   233.0  &    160  &  124   &  42.46  &  48.87 &  29.54  \\
\hline          
1676$^2$    & NGC 4568 & SAbc  &   357.0  &    160  &  124   &  42.46  &  48.87 &  29.54  \\  
\hline          
1690        &    M90   & SABab &   668.0  &    150  &   90   &  27.61  &  36.82 &  23.14  \\  
\hline          
1727        &    M58   & SABb  &   456.4  &    180  &  124   &  13.98  &  20.67 &  13.11  \\  
\hline
\end{tabular}
\end{center}
\end{table} 

An interesting case study is provided by close-by, interacting galaxies. Two such pairs are present in the proposed field, and are large enough to be studied in a spatially-resolved fashion.

The study of magnetic fields in interacting galaxies is primarily undertaken using synchrotron emission at radio wavelengths, hence tracing the magnetic fields associated with relativistic electrons. Interesting results have been found for the Antennae Galaxies (NGC 4038/4039). This well-studied interacting system shows strong magnetic fields concentrated along the tidal tails and the overlap region between the two galaxies. Observations with the Very Large Array (VLA) and the Effelsberg telescope by\cite{chizy04} revealed magnetic field strengths around 20-30 $\mu$G in these regions. The compression and turbulent motions in the overlapping region seem to enhance the magnetic field strength, which may play a role in triggering star formation. 

Magnetic field enhancement in the region of interaction in between the galaxy pair UGC 12914/UGC 12915 has also been reported by Ref. \cite{condon93}, who attributed this phenomenon to compression and turbulent effects during the collision. These observations seem to indicate that galaxy interactions can enhance and reorganize magnetic fields, especially in regions of strong gas compression and turbulence. Studying magnetic fields in these systems provides insight into how the former may be influenced by gravitational and hydrodynamical interactions. An interesting case can also be made considering the fact that these galaxies are immersed in the hot ICM, that can potentially alter these effects(e.g. Ref. \cite{draine79}).

Spatially resolved polarimetric observations are critical for advancing our understanding of dust grain properties and magnetic field structures across a wide variety of astrophysical environments. Unlike unresolved data, which average out local variations, spatially resolved polarimetry can reveal the intricate interplay between grain alignment, magnetic fields, and radiation fields on scales that, with the PPI instrument, vary from $\sim 800$ pc to $\sim 2$ kpc.

For example, variations in polarization across resolved regions of galaxies can indicate changes in grain alignment efficiency due to differing radiation field intensities or magnetic field orientations. In star-forming regions, spatially resolved data can disentangle contributions from dichroic extinction and scattering, identifying how the central radiation field interacts with the surrounding dust. This is essential for understanding dust grain growth, destruction, and alignment processes in environments ranging from the diffuse ISM to actively star-forming regions.

These kind of observations are also relevant for testing grain alignment theories, such as radiative alignment torques (RAT). Since the alignment efficiency of grains varies with radiation intensity and magnetic field orientation, the polarimetric SED can reveal how alignment changes across wavelengths, contributing to models of grain dynamics and alignment in environments dominated by intense radiation or magnetic fields\cite{lazarian07,andersson15}. This approach not only constrains the spatial variation of dust properties but also provides critical insights into the lifecycle of dust grains at galactic scales.

\section{Desired performances}
With the current design specifications for flux sensitivity (see: \url{https://prima.ipac.caltech.edu/page/instruments}, but see also the PRIMA Exposure Time Calculator: \url{https://prima.ipac.caltech.edu/page/etc-calc}), surveys aiming to map large areas, such as the one described in this work, can be carried out efficiently within a reasonable observing time.
The flux detection limits currently foreseen will enable the detection of very low dust masses, even at cold temperatures, particularly in local Universe galaxies. This opens the possibility to probe the faint, extended outskirts of galaxies that have so far proven to be challenging to study.
In Table~\ref{tab:mass_detect}, we report the minimum dust masses detectable with the PPI instrument under the assumption that dust emits as a modified black body with emissivity index $\beta = 1.8$, and opacity $\kappa_0 = 0.192$ m$^2$ kg$^{-1}$ at 350 $\mu$m, taken from\cite{draine03}. We provide estimates for dust temperatures of 15, 20, and 25 K, considering both point-like and extended sources. The beam is assumed to be circular, with FWHM values of 10.8, 14.8, 20.2, and 27.6 arcsec for the four PPI bands, respectively.

A similar estimation of dust mass limits for the PHI instrument is not straightforward. The PHI wavelength coverage (24–80 $\mu$m) is dominated by emission from stochastically heated small dust grains that are not in thermal equilibrium with the ambient radiation field. Their emission depends sensitively on several factors, including the intensity and spectral shape of the local radiation field, grain composition, and size distribution. As such, it cannot be reliably modeled with a modified blackbody function, which assumes thermal equilibrium, and which is the basis of the estimates presented here for PPI.

This further highlights a key complementarity between the two instruments: while PHI provides access to detailed information on grain properties and transient heating processes, PPI is uniquely suited for deriving bulk dust properties, including reliable dust mass estimates, through thermal continuum emission.

\begin{table}[t]
\caption{Dust mass detection limit, at each PHI band, for different dust temperature.}
\label{tab:mass_detect}
\begin{center}       
\begin{tabular}{|l|c|c|c|c|} 
\hline
          & PPI1 & PPI3 & PPI3 & PPI4  \\
\hline
\multicolumn{5}{|c|}{Point source mass detection [M$_\odot$]} \\
\hline
T = 15 K  & $2.10\times 10^5$ & $1.04\times 10^5$ & $7.95\times 10^4$ & $1.07\times 10^5$  \\ 
\hline
T = 20 K  & $1.72\times 10^4$ & $1.54\times 10^4$ & $1.95\times 10^4$ & $3.72\times 10^4$  \\ 
\hline
T = 25 K  & $3.84\times 10^3$ & $4.88\times 10^3$ & $8.27\times 10^3$ & $1.93\times 10^4$  \\  
\hline
\multicolumn{5}{|c|}{Extended source mass detection [M$_\odot$ beam$^{-1}$]} \\
\hline
T = 15 K  & $1.45\times 10^2$ & $8.03\times 10^1$ & $6.37\times 10^1$ & $8.48\times 10^1$  \\ 
\hline
T = 20 K  & $1.19\times 10^1$ & $1.19\times 10^1$ & $1.56\times 10^1$ & $2.96\times 10^1$  \\ 
\hline
T = 25 K  & $2.66           $ & $3.78           $ & $6.63           $ & $1.54\times 10^1$  \\ 
\hline
\end{tabular}
\end{center}
\end{table} 

Given these performance metrics, PRIMA is also poised to become the go-to instrument for polarization studies in the far-infrared, a regime traditionally limited by very low surface brightness levels.

\section{Conclusions}
Complementing {\it Herschel}’s FIR observations with data in the $24-90 \mu$m range, is needed to provide the constraints that will allow us to obtain critical insights into the warmer dust, star formation, and AGN activity in galaxies of the Virgo Cluster. These shorter wavelengths are perfectly suited to fill in key gaps in understanding the dust temperature distribution, mass estimates, and the dynamics of galaxy interactions. By capturing both the cold and warm dust components, PRIMA will offer a comprehensive view of the dust cycle, helping to address questions about galaxy evolution and cluster-specific processes. Furthermore, \textcolor{cyan}{this kind of observations can enable} not only to cover this previously missing range but, given the much better sensitivity of both of the PRIMA imagers, it \textcolor{cyan}{can} overcome the sensitivity limitation of PACS, which hampered the detection of several sources that were instead detected at SPIRE wavelengths.

The full access to the PRIMA capabilities and this wavelength range is also going to be essential to pin down the contributions of the different emission components (thermal dust emission, free-free, synchrotron, etc.) within galaxies. The Virgo cluster, with a combination of unique vicinity and environment characterization, is a ``must-do target'', where environmental effects on dust can be studied to the highest possible detail.

\subsection* {Acknowledgments}
The authors thank the two anonymous referees for their helpful comments and criticisms that helped improving the content ant clarity of this paper. JF acknowledges support from CONAHCyT, project number CF-2023-G100, and UNAM-DGAPA-PAPIIT IN111620 grant, Mexico.  E.F.-J.A. acknowledges support from UNAM-PAPIIT projects IA102023 and IA104725, and from CONAHCyT Ciencia de Frontera project ID: CF-2023-I-506.

This paper is dedicated to the memory of Jonathan Davies, P.I. of HeViCS and of DustPedia. 

\subsection* {Disclosures}
The authors declare there are no financial interests, commercial affiliations, or other potential conflicts of interest that have influenced the objectivity of this research or the writing of this paper.

\subsection* {Code and Data Availability}
Data sharing is not applicable to this article, as no new data were created or analyzed

\bibliography{article}   
\bibliographystyle{spiejour}   



\vspace{1ex}

\listoffigures
\listoftables

\end{spacing}
\end{document}